\documentclass[12pt]{iopart}

\usepackage{graphicx}
\usepackage{physics}
\usepackage{bbold}
\usepackage{color}
\usepackage{soul}
\usepackage{hyperref}

\usepackage{amsfonts}
\usepackage[normalem]{ulem}
\begin{document}

\title[Black hole information recovery from gravitational waves]{Black hole information recovery from gravitational waves}

\author{Louis Hamaide and Theo Torres}
\address{Department of Physics, King’s College London, The Strand, London WC2R 2LS, UK}
\ead{louis.hamaide@kcl.ac.uk}

\vspace{10pt}
\begin{indented}
\item[]November 2022
\end{indented}

\begin{abstract}
We study the classical and quantum black hole information in gravitational waves from a black hole's history. We review the necessary concepts regarding quantum information in many-body systems to motivate information retrieval and content in gravitational waves. We then show the first step in an optimal information retrieval strategy is to search for information in gravitational waves, compared to searching for correlations in Hawking radiation. We argue a large portion of the information of the initial collapsing state may be in the gravitational waves. Using the Zerilli equation for particles falling radially into Schwarzschild black holes, we then describe a method to retrieve full classical information about infalling sources, including masses, infall times and angles.

\end{abstract}

\setcounter{footnote}{0}

\section{Introduction}\label{sec:intro}
In response to the black hole information paradox, a vast body of literature discussing the problem in the fields of general relativity, string theory, holography and quantum information theory has appeared in the last 50 years (see~\cite{Harlow:2014yka,Raju:2020smc,Mathur:2009hf,Page:1993wv,Susskind:1993if,Horowitz:2003he,Hayden:2007cs,Almheiri:2012rt,Maldacena:2013xja,Strominger:2017zoo,Almheiri:2020cfm,Dai:2016jdv} and references therein for a non extensive overview). Assuming subtle correlations between evaporation quanta which are not taken into account in Hawking's initial calculation~\cite{PhysRevD.14.2460}, it is often expected that the final state can be made pure and information about the initial state retrieved (e.g.~\cite{Hayden:2007cs}). In doing so black holes are often implicitly portrayed as initially closed (or at least already equilibrated) thermal systems, allowing information to escape only in Hawking radiation. However since black holes are open systems throughout their history, they leak information at the early stages of their formation history, when gravitational waves are emitted. Further, although thermal many-body quantum systems and black hole perturbations have separately been extensively studied (see eg~\cite{Eisert:2014jea,doi:10.1146/annurev-conmatphys-031214-014726,DAlessio:2015qtq,Pound:2021qin,MaggioreMichele2,Pani:2013pma,Chandrasekhar1983} for a review), to our knowledge the two fields are quite separate and there has been no study of the quantum information content of gravitational waves from black holes. We will nevertheless find that a large portion of the information of an initial gravitationally collapsing state is present and easily accessible in gravitational waves. This is because a large share of the information of the state is stored in the classical observables of the state, which form a subset of the total information, usually referred to as quantum information\footnote{Note here that the term quantum information might be somewhat ambiguous as it refers to the total information contained in a system and therefore comprises the classical information as well as the purely quantum information. In this work we are interested in the share of classical information that can be retrieved.}, and which can be precisely recovered using black hole perturbation theory.

We believe that the gravitational wave community has therefore made extensive progress on accurately determining information from black holes in a way that is relevant to the information paradox, motivating further work to determine its content in GWs. 
In section 2 we review quantum information conservation in thermal systems and motivate information retrieval in GWs. In section 3, we review black hole perturbation theory in the case of a Schwarzschild black hole. In section 4 we show the algorithm for the inversion to obtain the masses of the infalling particle and black hole from the gravitational wave signal. Finally sections 5, 6 and 7 focus on obtaining the time and angle at which the particles fell in from the signal.

\section{Motivation \& introduction to black hole information}\label{sec:quantum_info_and_motivation}
\subsection{Concepts in density matrices, entanglement entropy and thermal systems}
Naively, black holes at rest are featureless and seem to destroy information as they evaporate. In the original paradox, Hawking showed the density matrix of the Hawking radiation was diagonal~\cite{PhysRevD.14.2460}, which he noted to be information-theoretically inconsistent with an initial ``pure" quantum state. Therefore before discussing gravitational waves, we begin by an introduction to quantum information conservation 
and motivate the study of black hole perturbations.

\begin{figure}[t!]
    \centering
    \includegraphics[width=.75\textwidth]{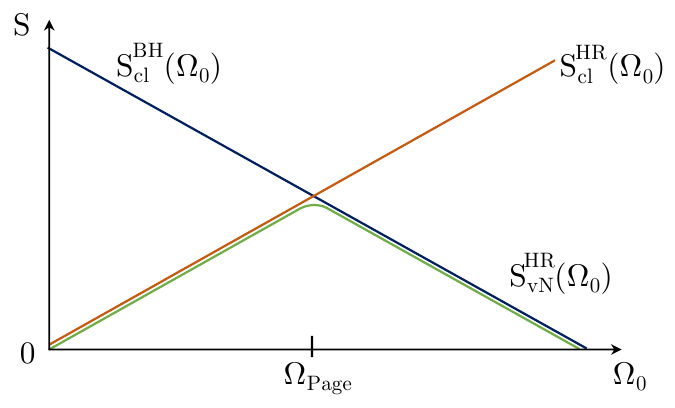}%
    \caption{We show in blue the evolution of the black hole thermodynamic (Bekenstein-Hawking) entropy derived from classical arguments $S_\textrm{cl}^\textrm{BH}=A/4$~\cite{PhysRevD.7.2333,Hawking1975}, in orange the classical radiation entropy $S_\textrm{cl}^\textrm{HR}$ from Hawking's calculation~\cite{Hawking1975}, and in green the Von Neuman entropy $S_\textrm{vN}(\Omega_0)=\Tr(\rho_{\Omega_0}\ln(\rho_{\Omega_0}))$ for a thermal pure state, as a function of the size of the subsystem $\Omega_0\subset\Omega$.}
    \label{fig:Hawking_vs_Page}
\end{figure}

Compared to wavefunctions, density matrices are generalized representations of states. States with well defined wavefunctions, called ``pure states", have $N\times N$ density matrices $\rho=\ket{\psi}\bra{\psi}$, where $\ket{\psi}=\sum_i a_i\ket{n_i}$ is the wavefunction in some basis $\ket{n_i}_{i\in[1,...,N]}$ and $N$ is the dimensionality of the Hilbert space. Off-diagonal elements $\rho_{i\neq j}$ can then appear as the phase coherence of the state. These phases are important as they guarantee the existence of a basis in which we can write $\rho=\mathbb{1}_{N\times N}\delta_{i,1}$. Pure states, however, are more often characterised using $\text{Tr}(\rho^2)=1$. More generally, a state with well defined probabilities but partial or absent phase coherence (e.g.~classical statistical ensembles) will exhibit a partial or total basis independent absence of off-diagonal matrix elements and is called a partially or totally ``mixed" state. Mixedness can be quantified as $\text{Tr}(\rho^2)<1$. Unitary evolution of the system, such as the time evolution of gravitationally collapsing radiation and/or matter, implies mixedness and vice versa purity remain constant since
\begin{align}
    \textrm{Tr}(\rho^2)=\textrm{Tr}(U\rho U^\dagger U\rho U^\dagger) \,\, .
\end{align}
This implies the mixedness/purity is a measure of information since information is assumed preserved in unitarily evolving systems. Further the Von Neuman entropy of a state, defined as $S_\textrm{vN}=-\textrm{Tr}(\rho \ln(\rho))$, is also a measure of mixedness of the system and remains constant. It is identically zero for a pure state only. But what of the classical entropy of a system described by a pure state which has thermalized? It is non zero as expected, and in fact can be found as the upper bound in the space of subsystems, which will be mixed, such that $S_\textrm{BH}(\Omega)=\max_{\Omega_0\subset\Omega}S_\textrm{vN}(\Omega_0)$. Indeed,~\cite{Page:1993df,Page:1993wv} showed that random subsystems from a thermal ensemble (such as blackbody radiation escaping a black hole) are mixed because they are entangled subsystems of a globally pure state. This is why Von Neuman entropy is sometimes called ``entanglement entropy". This can be illustrated with a simple two particle-two spin state example with states $\ket{\psi_\textrm{AB}}=\frac{1}{\sqrt{2}}(\ket{+-}-\ket{-+})$, taking the partial trace of the density matrix over the A states such that $\Tr_A(AB)=B\Tr(A)$ gives:
\begin{align}
    \rho_\textrm{AB}=\frac{1}{2}\left(\begin{matrix}
    0 & 0 & 0 & 0 \\
    0 & 1 & -1 & 0 \\
    0 & -1 & 1 & 0 \\
    0 & 0 & 0 & 0 \\
\end{matrix}\right)\enspace\longrightarrow\enspace
    \rho_\textrm{B}=\frac{1}{2}\left(\begin{matrix}
    1 & 0 \\
    0 & 1 \\
\end{matrix}\right)
\end{align}
\begin{align}
    S_\textrm{vN}(\rho_\textrm{AB})=0\enspace\longrightarrow \enspace S_\textrm{vN}(\rho_\textrm{B})=\ln 2 \,\, ,
\end{align}
where we expressed $\rho_\textrm{AB}$ in the basis $\{\ket{--},\ket{-+},\ket{+-},\ket{++}\}$, and summed over values (+ and -) of $\textrm{A}$ for each value of $\textrm{B}$ in the basis $\{\ket{+},\ket{-}\}$ to obtain $\rho_\textrm{B}$. Note this matches the classical entropy $S_\textrm{cl}(\textrm{AB})=\ln 2=S_\textrm{vN}(\rho_\textrm{B})$ since $\textrm{AB}$ has two possible microstates. Since the particles A and B are entangled, the density matrix of a subsystem can only contain partial information. Let us now assume a black hole is a thermalized body made of entangled quanta, which escape during evaporation. Since the global state is pure, as more particles escape the black hole the mixedness of the subsystem of evaporated quanta will eventually start to decrease after the ``Page time" until purity is restored. In the above example, assuming AB evaporates one particle at a time, the detector will initially see $S_\textrm{vN}(\rho_\textrm{vacuum})=0$, then $S_\textrm{vN}(\rho_\textrm{A or B})=\ln 2$ and finally $S_\textrm{vN}(\rho_\textrm{AB})=0$. This yield a so-called Page curve. However this is in contrast to Hawking's result which states the evaporation particles are maximally mixed, such that mixedness keeps increasing until the end of the evaporation, and emitting an entropy $S_\textrm{BH}$ (illustrated in figure \ref{fig:Hawking_vs_Page}).

\begin{figure}[t!]
    \centering
    \includegraphics[width=.8\textwidth]{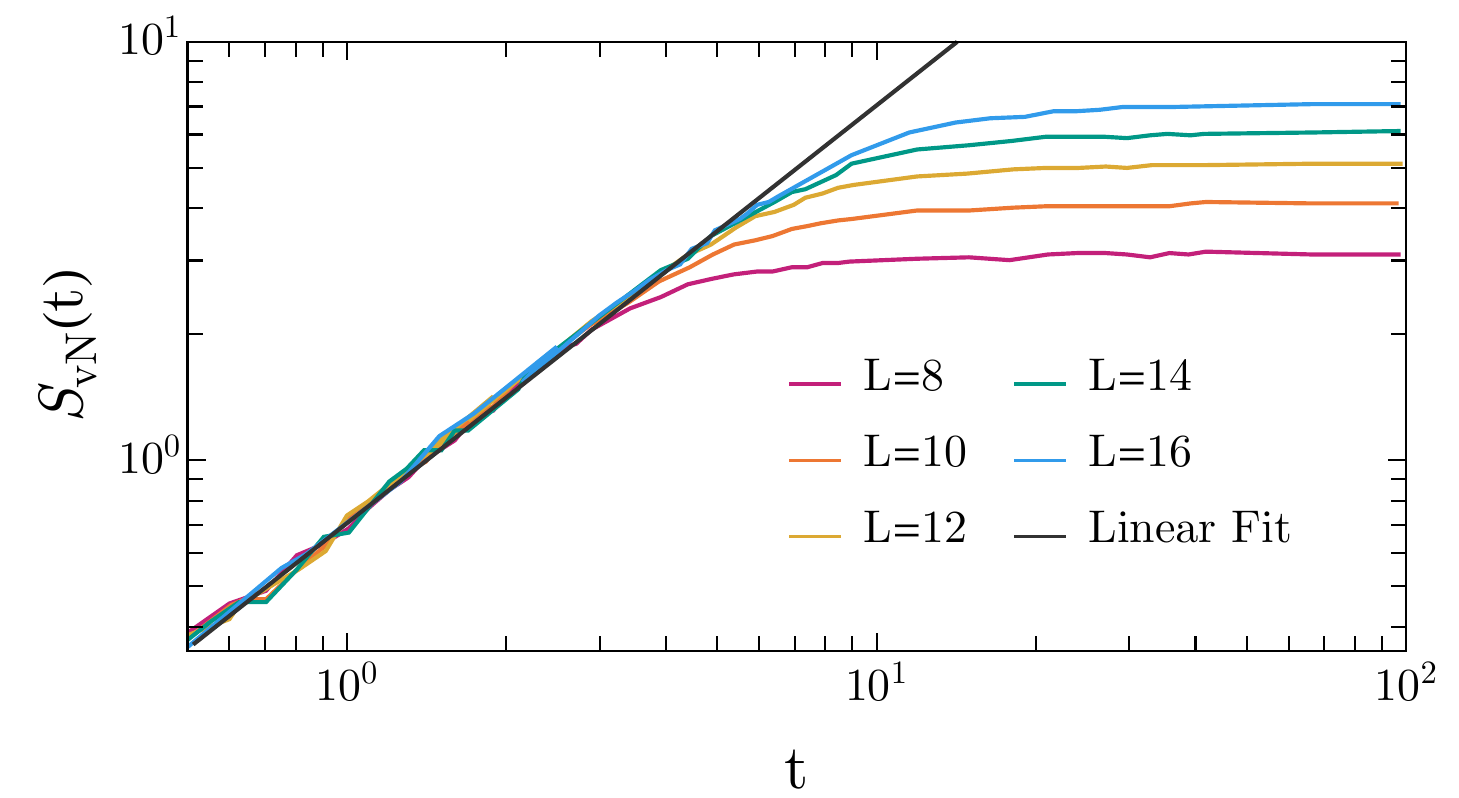}
    \caption{Evolution of the average entanglement entropy of half of a many-body system as a function of size L (data taken from ~\cite{PhysRevLett.111.127205}). $S_\textrm{vN}$ is a measure of the local correlations with the rest of the system. As the system thermalizes it converges up to small quantum fluctuations corresponding to the (small) interactions still occurring in the system, and which decrease relatively to the $S_\textrm{vN}$.}\label{fig:SvN_scrambling}
\end{figure}

Nevertheless recovering entanglement at late times has led the community to believe the information could be recovered if the black hole could be shown to evolve unitarily from a pure state and was referred to as the ``central dogma" of black hole information in~\cite{Almheiri:2020cfm}. However, this requires finding correlations between patches of the black hole metric, which could be transferred to the radiation. These correlations would naturally appear in a microscopic model of gravitational degrees of freedom that have become highly entangled through scatterings and eventually thermalizing. This also introduces along with it the idea of thermal metric fluctuations at the horizon, albeit small, as seen in figure \ref{fig:SvN_scrambling}. This also revisits the classical assumptions of a featureless background behind Hawking's initial calculation. Nevertheless, since the original formulation of the paradox, many have revisited the classical assumption of a purely classical the background metric (e.g.~\cite{PhysRevD.28.2929,Ford:1998he,Ford:1997zb,Barrabes:2000fr,Parentani:2000ts,Thompson:2008vi,Brustein:2013qma,PhysRevLett.114.111301,Dvali:2013lva,Dvali:2012rt,Gruending:2014rja,Calmet:2021stu,Calmet:2021cip,Casadio:2014vja,Corda:2021rly}).

These various works have described black hole horizons as a thermal microscopic ensemble in order to recover correlations. Assuming a pure state, the corresponding off-diagonal elements can be seen as quantum fluctuations \cite{DAlessio:2015qtq} and we will assume them to be present in rest of this work, although they have been subject of debate~\cite{Hawking1975,HAWKING1974}. Indeed, similarly to a classical state, a quantum many-body state also exhibits quantum fluctuations in addition to classical fluctuations at late times, when the system has ``equilibrated". This is defined as the moment the average local entanglement entropy has reached its maximum, or scrambling time, as illustrated in figure \ref{fig:SvN_scrambling}. The classical fluctuations can be seen as arising from the distribution of diagonal elements (representing probabilities) whereas quantum fluctuations can be seen as arising from the off-diagonal ones. These fluctuations appear due to interactions of the smaller microscopic degrees of freedom of the system. They can be used to recover the initial state in a closed system and therefore contain information. In the context of metrics, fluctuations are called ``hair", and have been studied extensively~\cite{Sotiriou:2013qea,Mavromatos:1995fc,Hawking:2016msc}. An important classical result in ``vanilla" general relativity is that asymptotically in time ``black holes have no hair"~\cite{1968CMaPh...8..245I,1971PhRvL..26..331C,1967PhRv..164.1776I}. Black hole hair appears therefore purely in models with quantum features.

As we have established quantum and classical fluctuations in an equilibrated system carry classical and quantum information (often referred to as ``coarse grained" and ``fine grained" respectively), we may study how to recover this information. The more entangled the final state, the more measurements and operations are necessary to recover quantum information (see e.g.~\cite{Vardhan:2021mdy,PhysRevA.92.042321}, although the algorithm may depend on the microscopic physics). After the scrambling time any algorithm to recover the information from fluctuations requires a larger number of steps, contrary to the initial state where the information is more readily available. Furthermore the size of fluctuations in the system reach their minimum after this time, and their size can be well modelled using the Eigenstate Thermalization Hypothesis (ETH), which has experimental and numerical success with a variety of generic systems~\cite{DAlessio:2015qtq}. The ETH describes local operators as
\begin{align}
    \hat{O}=f(\bar{E})\delta_{ij} + e^{-S_\textrm{cl}(\bar{E})/2}R_{ij}g(\bar{E},\omega) \,\, ,
\end{align}
where $(f(\bar{E}),g(\bar{E},\omega))$ are smooth functions of $(\bar{E},\omega)=((E_i-E_j)/2,E_i-E_j)$ and $R_{ij}$ is a matrix of random variables with zero mean and order one variance. Quantum fluctuations are therefore highly suppressed, and the classical fluctuations will dominate.

\subsection{Black hole quantum information in gravitational waves}
In our case, we are interested in fluctuations before the scrambling time, i.e.~at times near when a particle fell in, since they are larger and depend more heavily on the disentangled initial state $\ket{\text{infalling particle}}\otimes \ket{\text{black hole}}$. In the absence of a theory of quantum gravity, these can be calculated as gravitational waves using classical perturbation theory of a black hole metric with a source, which is exempt of the assumptions of the no-hair theorem\footnote{We will focus on $m_0\ll M_\textrm{BH}$ and use perturbation theory to allow us to show analytically that information is present as we see in the next section, although we believe these conclusions hold also in the strong gravity regime (eg black hole mergers).}. Corrections due to fluctuations of the metric from the scrambled past history of the black hole will still be present, but we may reasonably expect them to be smaller in a variety of scenarios. Indeed, assuming the thermodynamic limit of many microscopic gravitational degrees of freedom (quanta)
in thermal equilibrium with temperature $T_\mathrm{BH}=T_\mathrm{Hawking}$,
the energy fluctuations of the system can simply be estimated as the standard deviation of a Bose-Einstein condensate, which is $\mathcal{O}(kT_\mathrm{BH}=M_p^2/(8\pi M_\textrm{BH}))$. Meanwhile, the signal energy of early fluctuations of a radially infalling particle has been found to be $\int\abs{u(\omega,r)}^2d\omega dr 
\approx 10^{-2} m_0^2/M_\textrm{BH}\ll m_0$~\cite{PhysRevLett.27.1466,MaggioreMichele} where $u(\omega,r)$ is the GW signal while $m_0$ and $M_\textrm{BH}$ are the particle and black hole masses respectively (similar dependence on $m_0$ and $M_\textrm{BH}$ appear for more general trajectories). The recovered information will therefore have an SNR$^2(\propto\chi^2)\approx m_0^2/M_p^2$ where $M_p$ is the Planck mass.
Assuming the black hole and gravitational wave system is thermal such that information is spread evenly throughout the system, the quantum information recovered in this (sub)system can be quantified as the change in the Von Neuman entropy when taking the partial trace of the black hole density matrix over the subsystem such that:
\begin{align} \nonumber
    \rho_\textrm{BH$\otimes$GW}=\left(\begin{matrix}
     & \vdots &  \\
   \cdots & \ket{\textrm{BH}_i}\ket{\textrm{GW}_j}\bra{\textrm{GW}_k}\bra{\textrm{BH}_l} & \cdots  \\
    & \vdots & 
    \end{matrix}\right)
    \enspace \\ \longrightarrow\enspace 
    \rho_\textrm{BH}=\left(\begin{matrix}
     & \vdots &  \\
   \cdots & \ket{\textrm{BH}_n}\bra{\textrm{BH}_n} & \cdots  \\
    & \vdots & 
    \end{matrix}\right)
\end{align}
\begin{align}
    S_\textrm{vN}(\rho_\textrm{BH$\otimes$GW})=0 \enspace\longrightarrow\enspace S_\textrm{vN}(\rho_\textrm{BH})>0 \,\, ,
\end{align}
where $(i,j)$ and $(k,l)$ run over the Hilbert space of the black hole and the associated gravitational waves, in the columns and rows respectively, and whose states include all possible histories of the black hole given its initial conserved charges of mass, angular momentum, and possible gauge charges. Note this makes the simplest possible assumptions about $\rho_\textrm{BH$\otimes$GW}$. A more accurate estimate would require study of the thermalization and information loss of open quantum systems.

Further, although the information recovered by the gravitational wave spectrum will be recoverable up to small quantum fluctuations of the metric, the information in these fluctuations can also be retrieved in principle by measuring them.
Assuming Hawking radiation arises from the decay of microscopic gravitational degrees of freedom at the horizon, one expects it contains information that is approximately spread out through the horizon and
proportional to the size of the fluctuations which is given by the temperature. These fluctuations represent the late time tail of the black hole history \cite{Ford:1997zb,Barrabes:2000fr,Parentani:2000ts,Brustein:2013qma,Dvali:2012rt,Gruending:2014rja,Calmet:2021stu,Casadio:2014vja,Corda:2021rly,Maldacena:2001kr,Horowitz:1999jd}. 
We may therefore think of information retrieval in two parts: the first is reading large fluctuations in the form of gravitational waves from which information is easily retrievable, and the second is reading the late time tail fluctuations occurring after the scrambling time which encode the black hole history. Again note this tail of quantum and classical fluctuations appears as a consequence of the quantum features attributed to the horizon, since classical metric equations predict exponentially small signal with time in line with the no hair theorem. Further, the slow escape of the information from the horizon during the late time tail is in sharp contrast with early phase gravitational waves,
where the infalling particle is still mostly unentangled and coherently scatters with the microscopic degrees of freedom of the black hole, allowing for easy information recovery away from the horizon. 

\begin{figure}[t!]
    \centering
    \includegraphics[width=.75\textwidth]{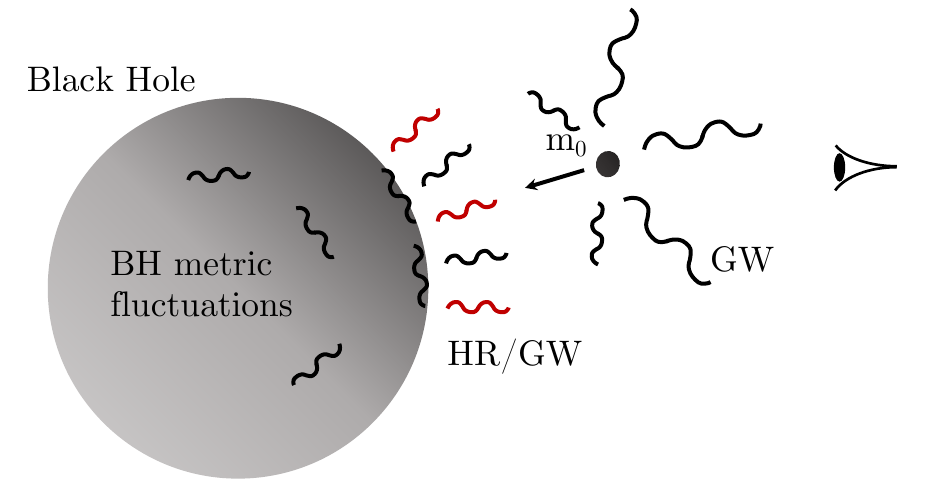}%
    \caption{Illustration of the information content of the metric. The black hole metric is assumed to be a thermal bath of microscopic degrees of freedom. As the bath evaporates through (red) blackbody Hawking quanta, the energy stored in the gravitational field is released, causing fluctuations (black) outside the horizon set by the blackbody temperature corresponding to noise in the measurement of the classical gravitational waves from the infalling particle. During the infall the system evolves from the product state $\ket{BH}_I\otimes\ket{m_0}_J$ to become part of the entangled thermal state $\sum_{i,j}\ket{BH}_i\ket{m_0}_j$.}
    \label{fig:BH_info_scrambling}
\end{figure}

We may also venture to estimate the share of the total information contained in the gravitational wave signal,
setting aside for now the difficulty of recovering the information. Assuming the Von Neumann entropy of a small subsystem grows linearly with the energy (which is an extensive quantity), 
we may estimate the information content of gravitational waves as their share of the total energy of the BH and GW system. This corresponds to taking the entanglement entropy of the black hole after formation but without the gravitational waves.
It can be expressed as the average share of energy $m_0$ emitted as GWs, which we call $E_\textrm{GW}$, during infall:
\begin{align}
    \int E_\textrm{GW}/m_0dt\approx \int 10^{-2}m_0/M_\textrm{BH}dt 
\end{align}
obtained over the black hole history, assuming most of the signal is recoverable (SNR $>1$).
However, this fails to account for the potentially large share of black hole history information present in the gravitational waves, which reduces/collapses the Hilbert space further. As we will see in section \ref{sec:inversion} this is $\mathcal{O}(1)$, such that the share of information of an observable can be estimated as $\mathcal{O}(1)\times\max(\textrm{SNR},1)$.
The total entropy in the system $S_\textrm{cl}$ is therefore reduced by a factor of $E_\textrm{fluct}/(E_\textrm{GW}+E_\textrm{fluct})= 1/(1+\textrm{SNR}^2)$, and
the fraction of the information in GWs becomes $1-1/(1+\textrm{SNR}^2)=\textrm{SNR}^2/(1+\textrm{SNR}^2)$,
which is large for information regarding any constituent of the black hole larger that the Planck mass. Thus gravitational waves contain the black hole information down to a Planck mass granularity, and the share of black hole information may be found as 
\begin{align}\label{eq:info_share}
\int_0^{M_\textrm{BH}}\textrm{SNR}^2(m_0)/(1+\textrm{SNR}^2(m_0))\text{d}m_0\approx 1-M_p/M_\textrm{BH} \,\,\, ,
\end{align}
which is quite large for astrophysical black holes.
Note we assume an $\mathcal{O}(1)$ share of the information is obtained for SNR$>1$, although a more conservative estimate would use SNR$>10$, and depends on the experimental setup \cite{thrane_talbot_2019}. However the main purpose of this work is to motivate information retrieval using gravitational waves and we leave this for future work.

The scope of black hole information retrieval in gravitational waves in different regimes is summarized in Table \ref{tab:info_summary}. In the classical picture ($\hbar\rightarrow 0$ such that $M_p\rightarrow 0$), black holes are eternal and information is lost behind the horizon. However the information of the initial state is also copied and reflected near the horizon in the form of gravitational waves, preserving the initial state information.
In the semiclassical picture, where fields are quantized on a smooth metric background, black holes evaporate into maximally mixed states, containing no information besides the conservation of total energy \cite{PhysRevD.14.2460}. In this context, gravitational waves provide the only source of information. Finally, in a purely quantum regime, one expects fluctuations of the black hole metric to contain information along with Hawking radiation. However by using only the GW signals (early fluctuations) to extract information we obtain the same share of information as in the semi-classical case (``quantum" picture in Table \ref{tab:info_summary}), i.e.~a sizeable portion of the initial state information, which converges to the classical result in the limit $M_p\rightarrow 0$.

\begin{table*}[t!]
\label{tab:summaryResults} 
\begin{tabular}{c c c c c c}
\multicolumn{1}{c}{  } & \multicolumn{1}{c}{\quad Information in GWs \quad } & \multicolumn{1}{c}{ Information in BH } & \multicolumn{1}{c}{ \quad Equations }  
\\ Classical picture & 1 & 0 & \quad \ref{eq:m0}, \ref{eq:degeneracy_period} \& \ref{eq:degeneracy_angle}
\\ Quantum picture & $\mathcal{O}(1-M_p/M_\textrm{BH})$ & \quad $\mathcal{O}(M_p/M_\textrm{BH})$ & \ref{eq:info_share}  \\  
\end{tabular}
\caption{Summary of share of information estimates of the initial collapsing state in GWs in both the fully classical picture (eternal black hole) and quantum case (including Hawking radiation and backreactions).}
\label{tab:info_summary}
\end{table*}

\subsection{Information recovery in the gravitational wave literature}
With this estimate we are left to quantify the other advantage of gravitational waves: the complexity, or relative simplicity, of recovering information from gravitational wave signals, which we do for black hole-particle mergers in the following sections. First however we comment on their role within the current body of literature on gravitational waves.

Gravitational wave signals from black hole perturbations have been studied extensively (e.g.~\cite{Pound:2021qin,MaggioreMichele2,PhysRevLett.27.1466}), and obtaining them to good numerical/computational accuracy is non trivial. However as we will see, once a few key quantities which can be computed in advance are known the dependence on the various quantum/classical numbers of the infalling particles and initial black hole is trivial, such that with full knowledge of the signal (i.e.~in the absence of noise) one can determine them exactly. This is in fact a common implicit assumption behind many gravitational wave detection works~\cite{thrane_talbot_2019,Ashton_2019,PhysRevD.91.042003,PhysRevD.93.024013}. However remarkably it has not been explicitly shown before, although mentioned in passing in~\cite{1973ApJ...185..649P}, partly due to its irrelevance in the face of traditional experimental constraints such as intrinsically noisy detectors, irreducible background signals and numerical accuracy. The information is therefore usually determined with a Bayesian approach using catalogues of waveforms (e.g.~\cite{PhysRevD.77.104017,1995PhRvD..52..605A,Babak:2017tow}) such that the widths of the significance contours go linearly with these uncertainties:
\begin{align}
P(\theta_S|d,\theta_N)=\frac{P(d|\theta_S,\theta_N)P(\theta_S|\theta_N)}{P(d|\theta_N)} \,\, ,
\end{align}
where $(\theta_S,\theta_N)$ represent vectors of parameters describing the signal and noise respectively and $d$ is the detector data. Nevertheless in an experimentally finite SNR regime, we may recast our previous statement regarding the accuracy of an observable by instead claiming that \textit{Bayesian contours do not give disjoint regions of high statistical significance when in the limit of highly sensitive/low noise experiments}.

\section{From Schwarzschild perturbations to GW data}\label{sec:review}
\subsection{Zerilli perturbation equation}
We will consider here a Schwarzschild black hole formed from particles radially falling in, perturbing the metric and emitting gravitational radiation. Other types of perturbation have been considered, and many are solvable to first order~\cite{PhysRevD.74.104020}. Schwarzschild black hole perturbations are separable into ten scalar degrees of freedom using a tensorial harmonic basis, which can be reduced to four polar (P-even) and two axial (P-odd) degrees of freedom by an appropriate choice of gauge~\cite{PhysRev.108.1063}. We may then solve the Einstein field equations of each set separately, resulting in the Zerilli and Regge-Wheeler equations~\cite{PhysRev.108.1063}. Furthermore Chandrasekhar proved these equations were equivalent (isospectral)~\cite{doi:10.1098/rspa.1975.0066} through a supersymmetry transformation. Chandrasekhar also proved~\cite{10.2307/79047} that the limit of the Teukolsky equations at zero spin, where Kerr spacetimes become Schwarzschild, is equivalent to the Regge-Wheeler and Zerilli equations, as expected. Thus we expect our method to be generalizable to Kerr perturbations. We also expect the method to generalize to Reissner-Nortstrom black holes as it reduces to a Schwarzschild metric in the limit of zero charge, and since perturbations on this background follow at Zerilli-like equation.

The Zerilli equation is obtained by perturbing a background Schwarzschild metric $g_{\mu\nu}=\textrm{diag}(-f(r),1/f(r),r^2,r^2\sin^2\theta)$, where $f(r)=(1-2M/r)$ and use $M=M_\textrm{BH}$ for notational brevity, with a small parity-even perturbation $h_{\mu\nu}^{even}$ such that $g_{\mu\nu}\rightarrow g_{\mu\nu}+h_{\mu\nu}^{even}$ and solving the field equations. Writing
\begin{equation}
    h_{\mu\nu}^{even}=
    \begin{pmatrix}
    -f(r)H_0Y_{lm} & -H_1Y_{lm} & 0 & 0 \\
    -H_1Y_{lm} & -\frac{1}{f(r)}H_2Y_{lm} & 0 & 0 \\
    0 & 0 & -r^2KY_{lm} & 0 \\
    0 & 0 & 0 & r^2\sin\theta K Y_{lm} \,\, , \\ 
    \end{pmatrix}
\end{equation}
where $Y_{lm}$ are spherical harmonics and $r_*=r+2M\ln(r/2M-1)$ is the tortoise coordinate\footnote{One reciprocally writes $2M(W(e^{r_*/2M-1})+1)$ where $W$ is the product logarithm.}, one can express the resulting equations in terms of just two out of the initial four scalar degrees of freedom, which are chosen to be $H_1$ and $K$ such that we discard $H_0$ \& $H_2$. With the following choice of variable
\begin{align}\label{eq:uhat_dof} \nonumber
    \hat{u}=\alpha K+i\beta H_1
\end{align}
\begin{align}
    \alpha=\frac{r^2}{\eta r+3M} \,\,\,\,\,,\,\,\,\,\, \beta=\frac{2M-r}{\omega(\eta r+3M)}
\end{align}
one can write the radial Zerilli equation ~\cite{PhysRevD.2.2141}:
\begin{equation}\label{eq:Zerilli}
    \frac{d^2\hat{u}(\omega,r_*)}{dr_*^2}+(\omega^2-V(r_*))\hat{u}(\omega,r_*)=I(\omega,r_*) \,\, ,
\end{equation}
where
\begin{equation}
    V(r)=\left(1-\frac{2M}{r}\right)\frac{2\eta^2(\eta+1)r^3+6\eta^2Mr^2+18\eta M^2r+18M^3}{r^3(\eta r+3M)^2}
\end{equation}
with $\eta=(l-1)(l+2)/2$ and as usual $(M,\omega,l)$ are the black hole mass, frequency (eigenvalue of the Zerilli equation) and spherical harmonic $l$-number. 
Therefore as expected from the e.g.~de Donder gauge, we see perturbations of a Schwarzschild metric (gravitational waves) are solutions of a wave equation with admit two real degrees  of freedom: polarizations $K$ and $H_1$. In what follows we will assume that there is a one-to-one mapping between experimental GW data, $H_1$ and $K$, and $\hat{u}$. We will further assume a \textit{gedanken} experiment which allows us to obtain GW signals and $\hat{u}$ from any point $x^\mu$ and with arbitrary accuracy.

\subsection{Source term}
The right hand side of \ref{eq:Zerilli} describes the source of the perturbations:
\begin{align}\label{eq:source}
    I(\omega,r)=\frac{4m_0e^{i\omega T(r)}}{\eta r+3M}\left(l+\frac{1}{2}\right)^{1/2}\left(1-\frac{2M}{r}\right)\left[\left(\frac{r}{2M}\right)^{1/2}-\frac{2i\eta}{\omega(\eta r+3M)}\right] \\ \nonumber + e^{i\omega t_0}\left[i\omega u(t,r)-\pdv{u(t,r)}{t}\right]_{t=t_0} \,\, ,
\end{align}
where $m_0$ is the infalling particle mass $m_0\ll M$. The first sum term of $I$ is called the ``source term" denoted $\mathcal{S}$ whereas the second term represents perturbations present at $t_0$ when we start observing GWs, and appears in the frequency domain as an artifact of the Laplace transformation of the time domain Zerilli equation:
\begin{equation}
    \frac{\partial^2u}{\partial r_*^2}-\frac{\partial^2u}{\partial t^2}+V(r_*)u=\mathcal{S} \,\, ,
\end{equation}
where the Laplace transform is:
\begin{equation}
    \mathcal{L}u(t,r_*)=\hat{u}(\omega,r_*)=\int_{t_0}^\infty u(t,r_*)e^{i\omega t}dt \,\, .
\end{equation}
For now we may ignore the second term in $I(\omega,r)$ if the signal is detected such that $u\approx 0$ for $t_0\rightarrow -\infty$. $T(r)$ describes the coordinate time at which the particle can be observed at radius $r$, which for radial infall from rest is~\cite{PhysRevLett.27.1466}
\begin{equation}
    \frac{T(r)}{M}=-\frac{4}{3}\left(\frac{r}{2M}\right)^{3/2}-4\left(\frac{r}{2M}\right)^{1/2}+2\log\left[\left(\left(\frac{r}{2M}\right)^{1/2}+1\right)\left(\left(\frac{r}{2M}\right)^{1/2}-1\right)^{-1}\right] \,\, .
\end{equation}
While the expression for $T(r)$ and hence $I(\omega,r)$ appear cumbersome, remarkably $m_0$ only appears as an overall factor in $I(\omega,r)$, which allows us to extract information about the falling mass relatively easily, as we will see in the next section.

\subsection{Solution to the wave equation}
One may solve the Zerilli equation \ref{eq:Zerilli} using a Green's function 
such that:
\begin{equation}\label{eq:greens_fct}
    G(\omega,r_*,\zeta)=\frac{1}{W(\zeta)}
        \begin{cases}
        \hat{u}_1(\omega,r_*)\hat{u}_2(\omega,\zeta) \,\,\text{ if }r_*\leq \zeta\\
        \hat{u}_1(\omega,\zeta)\hat{u}_2(\omega,r_*) \,\,\text{ if }r_*\geq \zeta \,\, ,
        \end{cases}
\end{equation}
where $\hat{u}_1$ (respectively $\hat{u}_2$) is a solution to the homogeneous Zerilli equation, i.e.~$I=0$, that respect the desired boundary condition at the horizon (respectively infinity) and $W=\pdv{\hat{u}_1}{r_*}\hat{u}_2-\pdv{\hat{u}_2}{r_*}\hat{u}_1$. The solution to \ref{eq:Zerilli} is then:
\begin{equation}\label{eq:greens_int}
    \hat{u}(\omega,r_*)=\int_{-\infty}^\infty I(\omega,\zeta)G(\omega,r_*,\zeta)d\zeta \,\, .
\end{equation}
The Green's function is a solution to the Zerilli equation with a point-like source function $I$ located at $\zeta$. Since solutions to the Zerilli equation are linear in the perturbations, one may sum over a distribution of point-like potentials. The solution to \ref{eq:greens_int} is then the weighting of the Green's function against a distribution $I$, which is a plane wave with time dependent phase for radially infalling point particles.

More specifically, the $r_*\rightarrow\pm\infty$ boundary conditions on the $\hat{u}_i(\omega,r_*)$ solutions allow us to write in a convenient plane wave basis:
\begin{equation}\label{eq:u1asymptotic}
    \lim_{r_*\rightarrow -\infty}\hat{u}_1(\omega,r_*)=e^{-i\omega r_*}
    \,\,\,\,\,\,\text{ , }\,\,\,\,\,\,
    \lim_{r_*\rightarrow +\infty}\hat{u}_1(\omega,r_*)=A_{out}(\omega)e^{i\omega r_*}+A_{in}(\omega)e^{-i\omega r_*}
\end{equation}
and
\begin{equation}\label{eq:u2asymptotic}
    \lim_{r_*\rightarrow +\infty}\hat{u}_{2\pm}(\omega,r_*)=e^{\pm i\omega r_*}
    \,\,\,\,\,\,\text{ , }\,\,\,\,\,\,
    \lim_{r_*\rightarrow -\infty}\hat{u}_{2\pm}(\omega,r_*)=B_{out}(\pm\omega)e^{\pm i\omega r_*}+B_{in}(\pm \omega)e^{\mp i\omega r_*} \,\, ,
\end{equation}
We define quasinormal modes (QNMs) $\omega_{nl}$ as having $A_{in}(\omega_{nl})=0$. Note that we define two $\hat{u}_2$ functions for later use but will only use $\hat{u}_2=\hat{u}_{2+}$ in the Green's function since we assume no incoming radiation from infinity\footnote{Also note normalizing the solutions and taking the $r_*\rightarrow -\infty$ limit gives $B_{out}(\omega)=A_{in}(\omega)$ and $B_{in}(\omega)=-A_{out}(-\omega)=-A_{out}^*(\omega)$~\cite{PhysRevD.34.384}.}. From \ref{eq:u1asymptotic} \& \ref{eq:u2asymptotic} we find that $W=2i\omega A_{in}(\omega)$. We can now write the solution in $\omega$-space:
\begin{equation}\label{eq:solution_full}
    \hat{u}(\omega,r_*)=\frac{\hat{u}_2(\omega,r_*)}{2i\omega A_{in}(\omega)}\int_{-\infty}^{r_*}I(\omega,\zeta)\hat{u}_1(\omega,\zeta)d\zeta + \frac{\hat{u}_1(\omega,r_*)}{2i\omega A_{in}(\omega)}\int_{r_*}^{+\infty}I(\omega,\zeta)\hat{u}_2(\omega,\zeta)d\zeta \,\, .
\end{equation}
If the source has compact support near the horizon 
and the experiment, located at $r_*$, is far from the horizon and the source, this simplifies to:
\begin{equation}\label{eq:solution}
    \hat{u}(\omega,r_*)= \frac{e^{i\omega r_*}}{2i\omega A_{in}(\omega)}\int_{-\infty}^{+\infty}I(\omega,\zeta)\hat{u}_1(\omega,\zeta)d\zeta \,\, .
\end{equation}
This may further simplify if the source is sufficiently close to the horizon to replace $\hat{u}_1\sim e^{-i\omega\zeta}$.
Although this may not be our case, in what follows we will focus on this integral for simplicity, since the result and techniques can be straightforwardly generalized to the case of \ref{eq:solution_full}. At first glance the integral in \ref{eq:solution} is typically divergent at the horizon. Nevertheless we may regularize it by imposing appropriate boundary conditions (see Appendix).

\begin{figure}[t!]
    \centering
    \includegraphics[width=.75\textwidth]{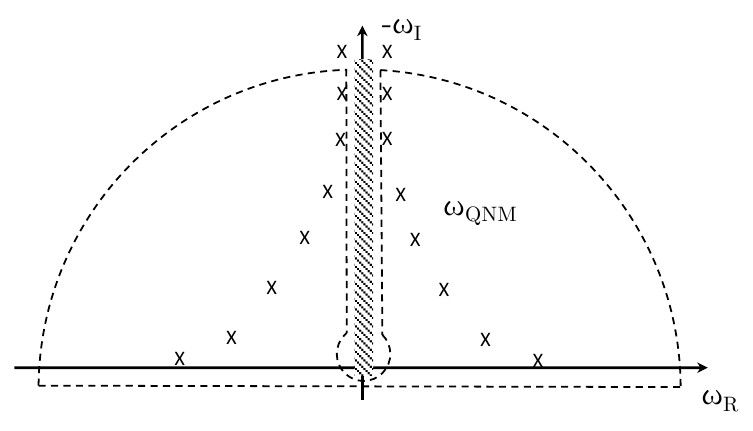}%
    \caption{Schematic contour of integration for the inverse Laplace transform of \ref{eq:solution}. Often a branch cut (hashed region) is introduced for values with $\omega_R=0$ which don't oscillate and thus aren't QNMs (non-radiative), however we include them in our work without loss of generality (\ref{eq:signal}). Their physical significance still requires careful study, but may represent higher order contributions to the signal or late time tail effects, since both contribute late-time tails (see e.g.~\cite{PhysRevD.74.104020} and references therein).}\label{fig:QNM_contour}
\end{figure}

We now want to understand the relation with the time series which is measured at our \textit{gedanken} experiment. We must therefore return to the time domain via an inverse Laplace transform. This integral is non-trivial and requires defining a contour in the complex $\omega$-plane (see figure \ref{fig:QNM_contour}). The contour contains poles at the quasinormal frequencies $\omega_{nl}$ where $A_{in}(\omega_{nl})=0$, which following the residue theorem contribute to the result:
\begin{align}\label{eq:signal} \nonumber
    u(t,r_*)=\sum_{n,l}\frac{A_{out}(\omega_{nl})}{2\omega}\left(\frac{dA_{in}(\omega_{nl})}{d\omega}\right)^{-1}e^{-i\omega_{nl}(t-r_*)}\int_{-\infty}^{+\infty}\frac{I(\omega_{nl},\zeta)\hat{u}_1(\omega_{nl},\zeta)}{A_{out}(\omega_{nl})}d\zeta  \\
    -  \frac{1}{2\pi}\int_{HC+BC}\left( e^{-i\omega(t-r_*)}\int_{-\infty}^{+\infty}I(\omega,\zeta)\hat{u}_1(\omega,\zeta)d\zeta\right) d\omega \,\, .
\end{align}
Note the inverse derivative of $A_{in}$ is an intrinsic property of the black hole metric, and combined with the source integral is often called ``quasinormal excitation coefficient" (QNEC). The observed radiation time series is thus a sum over QNMs $\omega_{nl}$ and a continuous spectrum of half circle modes and the branch cut. We discuss in the following sections the effect of the contour on information recovery.

\subsection{Finding the QNMs and $A_{in}$'s}

Although the $A_{out}$'s cancel, the residue coefficient requires knowledge of $A_{in}$ in the neighborhood of $\omega_{nl}$. We follow~\cite{PhysRevD.74.104020,PhysRevD.34.384} in determining these using the following series ansatz in $\omega$-space and Schwarzschild coordinates with $2M=1$\footnote{A simpler analytical approach is possible from~\cite{PhysRevD.74.104020} in the limit of large $n$.}:
\begin{align}\label{eq:u_series}
    \hat{u}_1(\omega,r) & =(r-1)^{-i\omega}r^{2i\omega}e^{i\omega(r-1)}\sum_{j=0}^\infty a_j\left(\frac{r-1}{r}\right)^j \\ 
    \hat{u}_{2\pm}(\omega,r) & =(2\omega)^{\mp i\omega}e^{i\phi_\pm}(1-\frac{1}{r})^{-i\omega}\sum_{j=-\infty}^{+\infty}b_j(G_{j+\nu}(-\omega,\omega r)\pm iF_{j+\nu}(-\omega,\omega r)) \,\, , 
\end{align}
where $G_{j+\nu}$ and $F_{j+\nu}$ are the regular and irregular Coulomb wave functions with $\nu$ chosen to yield a minimal solution for the series $b_j$ (i.e.~with a fixed initial value $a_0$, which \ref{eq:u1asymptotic} fixes to $b_0=1$) and 
\begin{align}
    \phi_\pm=\pm i\ln\left[\sum_{j=-\infty}^{+\infty}b_j\left(\frac{\Gamma(j+\nu+1-i\omega)}{\Gamma(j+\nu+1+i\omega)}\right)^{\pm 1/2}e^{\mp i(j+\nu)\pi/2}\right] \,\, .
\end{align}
These expansions respect the correct boundary conditions e.g.~
\begin{equation}\label{eq:u1series}
    \hat{u}_1(\omega,r)\sim (r-1)^{-i\omega}e^{i\omega(r-1)} \sim e^{-i\omega r_*} \,\,\,\,\,\,\,\text{ as }\,\,\, r\rightarrow 1 \,\, ,
\end{equation}
Although the above expansion holds for all $r_*$, the series $a_j$ diverges as $r\rightarrow\infty$, and is therefore impractical for direct evaluation of $A_{out}(\omega_{nl})$.
In \ref{eq:u1asymptotic} \& \ref{eq:u2asymptotic}, we have defined three solution to the homogeneous Zerilli equation, namely $u_1$, $u_{2-}$ and $u_{2+}$, each respecting a different boundary condition. Any pair of these three solutions will form a basis of the space of solutions to the homogeneous Zerilli equation, which is a second order linear ODE. Hence, any one of these solutions can be expressed as a linear superposition of the other two (the combination is oftentimes referred to as a Bogoliubov transformation). The coefficient of the Bogoliubov transformation can be found from the boundary conditions that defines the various solutions, in particular we can write
\begin{equation}
    \begin{cases}
    \hat{u}_1(\omega,r_*)=A_{in}(\omega)\hat{u}_{2-}(\omega,r_*)+A_{out}(\omega)\hat{u}_{2+}(\omega,r_*) \\
    \hat{u}_1'(\omega,r_*)=A_{in}(\omega)\hat{u}_{2-}'(\omega,r_*)+A_{out}(\omega)\hat{u}_{2+}'(\omega,r_*) \,\, .
    \end{cases}
\end{equation}
Thus we can solve for $A_{out}$ and $A_{in}$ for all $\omega$ and some chosen finite $r_*$\footnote{This method is now less common for computational reasons, in favor of~\cite{Mano:1996mf}.}:
\begin{equation}\label{eq:Ain_inversion}
    \begin{cases}
    A_{in}(\omega)= \frac{u_1(\omega,r_*)u_{2+}'(\omega,r_*)-u_1'(\omega,r_*)u_{2+}(\omega,r_*)}{u_{2+}'(\omega,r_*)u_{2-}(\omega,r_*)-u_{2+}(\omega,r_*)u_{2-}'(\omega,r_*)} \\ 
    A_{out}(\omega)= \frac{u_1'(\omega,r_*)u_{2-}(\omega,r_*)-u_1(\omega,r_*)u_{2-}'(\omega,r_*)}{u_{2+}'(\omega,r_*)u_{2-}(\omega,r_*)-u_{2+}(\omega,r_*)u_{2-}'(\omega,r_*)} \,\, ,
    \end{cases}
\end{equation}
from which we can calculate $dA_{in}/d\omega_{nl}$ once we obtain $(\hat{u}_1,\hat{u}_{2\pm})$. To do this we plug the series into the Zerilli equation. 
The recursion relation obeyed by the coefficient of the series is given by (re-establishing mass units where $2M\neq 1$):
\begin{align}\label{eq:recursion} \nonumber
    \alpha_0 a_1+\beta_0 a_0=0 \\
    \alpha_{nl}a_{n+1}+\beta_{nl}a_n+\gamma_{nl}a_{n-1}=0
\end{align}
\begin{equation}\label{eq:alpha_beta_gamma_coefs}
\begin{split}
    \alpha_{nl}= & \, n^2+(2-4i\omega M)n+1-4i\omega M \\
    \beta_{nl}= & -(2n^2+(2-16i\omega M)n-32\omega^2 M^2-8i\omega M+l(l+1)-3) \\
    \gamma_{nl}= & \, n^2-8i\omega M n-16\omega^2 M^2-4 \,\, ,
\end{split}
\end{equation}
where $n$ corresponds to the series index.
Building on the three term recurrence relation, Leaver showed that the solution to the Zerilli equation will converge~\cite{doi:10.1098/rspa.1985.0119,PhysRevD.34.384,leaver1986solutions} if the $(a_n)_{n\in\mathbb{N}}$ is a minimal solution to the above recurrence relation. The existence of a minimal solution implies that the following continued fraction holds:
\begin{equation}\label{eq:CFM}
    \frac{a_1}{a_0} = \frac{-\gamma_{1l}}{\beta_{1l}-}\frac{\gamma_{2l}\alpha_{1l}}{\beta_{2l}-}\frac{\gamma_{3l}\alpha_{2l}}{\beta_{3l}-}... \quad ,
\end{equation}
where we have used the standard notation for continued fraction in the above equation.
\ref{eq:CFM} will be valid for a discrete set of complex frequencies which are the QNM frequencies. The index labeling the element of this set is called the ``overtone'' number. In practice one numerically solves \ref{eq:CFM} or its $n^{\text{th}}$ inversion to find the $n^{\text{th}}$ QNM frequencies. This method of finding QNM frequencies, commonly called Leaver's method has been improved and generalised to a wide range of space-time beyond the Schwarzschild metric considered here.
QNM solutions ($A_{in}(\omega)=0$) occur at fixed discrete solutions of $u_i(\omega,r_*)$ and $\{\alpha_{nl},\beta_{nl},\gamma_{nl}\}_{n\in\mathbb{N}}$. By inspection of \ref{eq:alpha_beta_gamma_coefs}, the coefficients remain constant if $\omega_{nl}$ is simply related to the mass:
\begin{equation}\label{eq:QNM_propto_MBH}
    \omega_R\propto M_\textrm{BH}^{-1} \,\,\, , \,\,\,\,\,\, \omega_I\propto M_\textrm{BH}^{-1} \,\,\, ,
\end{equation}
The proportionality coefficients are determined solely by the structure of the black hole perturbation equations and $(n,l)$ and can be determined up to arbitrary accuracy (i.e.~given sufficient computational resources) by inverting the continued fraction. For the ``fundamental" frequency $(n,l)=(0,2)$ they are often quoted as $(12.07,18.06)$~kHz~$M_\odot$. Having explained how to calculate each of the quantities of \ref{eq:signal} our \textit{gedanken} GW experiment may find the mass of the black hole and the rest of the parameters with arbitrary accuracy, as we will see in the following section.

\section{Inversion}\label{sec:inversion}
We are now ready to invert the GW signal to retrieve physical information \textit{(semi-) analytically}. As mentioned previously, the GW signal will contain information regarding the initial black hole and perturbing infalling mass. To our knowledge the method for extracting the information in the case of a \textit{gedanken}  experiment with arbitrarily high SNR and spatial access to the signal outside the horizon has not been shown explicitly before. In practical cases the detector will be noisy and cannot be easily placed at several points around the black hole or take data for arbitrarily long times with arbitrarily high spatial and temporal resolution. The signal noise will then propagate to the amplitudes in the Laplace transform, and subsequent parameters deduced from them, and justifies instead the Bayesian approach used by gravitational wave astronomers. However it is of classical information-theoretical importance to know whether there exist unresolvable degeneracies in the signal, which next generation GW detectors might encounter in the near future. 

To invert the signal we assume in this subsection no a priori knowledge of the mass of the black hole or the perturbing particle, other than the assumption of the black hole being Schwarzschild, although generalization to Kerr metrics should be possible and is reserved for future work. We wish to reconstruct both $m_0$ and $M_\textrm{BH}$ to arbitrary precision, assuming arbitrary precision of our GW data can first be obtained (i.e.~SNR$\rightarrow\infty$).

Before $t=t_0$ we must obtain $r_*$, the location of the detector with respect to the singularity, as it will soon become useful. During this time one can match the readings of two or more detectors
around the black hole whose measurements are found to coincide when rotated, assuming each detector is of finite non-zero size and orientable. By taking lines aligned with these detectors and which pass through the detectors' centers one finds the singularity as the point the lines intersect. Alternatively, for idealized point-like non-orientable detectors we may find the distance to the singularity by finding the equation of sphere of four points with the same signal using e.g.~\cite{Beyer1987}.

When measuring the signal at $t\geq t_0$, assuming the signal comes from a mass infalling at angle $(\theta,\phi)$ and that the GW detector acts on a sphere of arbitrary radius which we assume near the horizon, this angle must first be found (as shown in Section \ref{sec:angle_recovery}) to divide out the spherical harmonic from which the radial component $u(r_*,t)$ can be deduced. The radial signal $u(r_*,t)$ comes in the form of \ref{eq:signal}, where we have measured the half circle modes as well. We then may Laplace transform the signal to obtain the frequency space representation $\hat{u}(\omega,r_*)$ \ref{eq:solution}.

Next, we observe the $\omega$'s at which the divergence of \ref{eq:solution} occurs: these are our QNMs frequencies $\omega_{nl}$. The frequency with the lowest imaginary part $\omega_I$, or ``fundamental'' frequency, immediately yields the black hole mass, as discussed around \ref{eq:QNM_propto_MBH}. To get the mass $m_0$, we first remove the radial phase $e^{i\omega_{nl}r_*}$ in the complex mode amplitudes with knowledge of the detector's distance from the horizon $r_*$ and the QNM values $\omega_{nl}$. With the black hole mass in hand, we may also calculate the coefficients $(\alpha_{nl},\beta_{nl},\gamma_{nl})$, which depend on $M_\textrm{BH}$ once physical units are reintroduced,\footnote{Note here we are talking about $(\alpha_{nl},\beta_{nl},\gamma_{nl})$ for \textit{all} $\omega$, not just minimal solutions.} and in turn $(\hat{u}_1(\omega,r_*),\hat{u}_2(\omega,r_*))$ according to \ref{eq:u_series} \& \ref{eq:recursion}. Using $(\hat{u}_1(\omega,r_*),\hat{u}_2(\omega,r_*))$ we may then determine $A_{in}(\omega)$ and $dA_{in}/d\omega_{nl}$ using \ref{eq:Ain_inversion}.

We remind ourselves of the expression for the source, i.e.~\ref{eq:source} and ignoring the boundary term. One may now perform the integral in \ref{eq:solution}:
\begin{equation}\label{eq:sourceint}
    \int_{-\infty}^{+\infty}I(\omega,\zeta)\hat{u}_1(\omega,\zeta)d\zeta
\end{equation}
using the regularization technique described in the Appendix, such that it converges for all $\omega$. The source term being simply proportional to $m_0$, the mass of the infalling particle we wish to recover, we may write:
\begin{equation}\label{eq:m0}
    m_0=\frac{2\omega A_{in}(\omega)e^{-i\omega r_*}\hat{u}(\omega,r_*)}{\int_{-\infty}^{+\infty}I_0(\omega,\zeta)\hat{u}_1(\omega,\zeta)d\zeta} \,\, ,
\end{equation}
where $I_0=I/m_0$. Note that this equation is valid for QNMs as a limit $\omega\rightarrow\omega_{nl}$. We require QNMs as the series $u_i(\omega,r_*)$ do not converge or respect the boundary conditions at other frequencies~\cite{doi:10.1098/rspa.1985.0119}. 

We have now peeled away each layer of \ref{eq:signal} semi-analytically, revealing the quantities $(M_\textrm{BH},m_0)$ determined to arbitrary accuracy assuming the GW signal from the particle's infall could be detected with arbitrary accuracy. Quantities which don't depend on $m_0$ such as $A_{in}(\omega)$ or \ref{eq:sourceint} may be evaluated numerically using $M_\textrm{BH}$ found from the signal $\hat{u}(\omega,r_*)$. As we will see in the following section we may also extract the time and angle of infall.

\section{Time recovery}\label{sec:time_recovery}
In this section we briefly discuss determining the time of the infall which we will call $t_0$. Ignoring any angular momentum in the history of the black hole means the classical information should consist entirely of the masses, times, and angles from which the particles radially fell in. Having seen in the previous section how to obtain the masses $m_0$ and $M_\textrm{BH}$, we turn our attention to the time. 

Infall time $t_0$ may be defined with reference to any point during the infall, since the observer never sees the particle fall in and we are not provided a natural reference time $t_0$. However we may choose an arbitrary radial coordinate $r_{*0}$ at which we wish to know the time $t_0$ when the particle crossed by looking only at the GW signal. We start by assuming a Schwarzschild reference frame with an arbitrary moment $t=0$, where $t$ is the coordinate time. $t_0$ is then the coordinate time at which the particle is considered to fall, or be dropped, into to the black hole. Following \ref{eq:greens_int} we assume a radial infall from infinity with zero energy.
We assume without loss of generality the observation window includes a period where the particle has been en route to the black hole for enough time to start creating a GW signal larger than the arbitrarily small (but larger than zero) noise of the experiment, which we may choose to be at the level of the metric fluctuations. Although we do not need to observe the entire signal to obtain information, we should always try to include the period of largest GW signal in the observation window such that the quantity/quality of information is optimized.

The GW spectrum in $\omega$-space is characterised by \ref{eq:solution}. We recall that the only difference in the particle infalling earlier or later is an overall time translation, which we call $t_0$, such that
\begin{align}
    \text{``earlier''} & \rightarrow e^{i\omega_{nl}(t-r_*)} \\
    \text{``later''} & \rightarrow e^{i\omega_{nl}(t+t_0-r_*)} \,\, .
\end{align}
This $e^{i\omega_{nl} t_0}$ phase is carried over to $\hat{u}(\omega,r_*)$, notably in \ref{eq:m0} where we had implicitly assumed $t_0=0$. Generally this introduces a phase, which may be recovered by finding $t_0$ is some frequency up to a periodic degeneracy
\begin{align}\label{eq:degenacy_period_onemode}
    A_{nl}(\omega_{nl})=e^{i\omega_{nl} t_0}\iff t_0=t_0+ 2\pi n/\omega_{nl} \,\, ,
\end{align}
where $A_{nl}$ represents the computed quantity \ref{eq:m0} and $n\in\mathbb{Z}$, and which is bounded above by the inverse of the experiment's observation period $2\pi/\omega_{min}$. We may however hope improve the degeneracy by obtaining $A_{nl}$ for two frequencies and observing that
\begin{align}\label{eq:degeneracy_period}
    t_0+2\pi n/\omega_0=t_0+2\pi m/\omega_1\iff n=\frac{\omega_1}{\omega_0}m 
\end{align}
is solved when $(m\,\omega_1)/\omega_0$ is an integer, which may require $(n,m)\gg 1$ or may have only one solution $m=n=0$ depending on $(\omega_0,\omega_1)$ and would either reduce the degeneracy by increasing it to larger periods $(n,m)\times 2\pi/\omega$, or remove it entirely by fixing $(n,m)$ to zero. In a gedanken experiment we may assume the observation time is infinite,
thus avoiding the degeneracy in \ref{eq:degenacy_period_onemode}. However since we are not guaranteed QNMs with arbitrary small $\omega_R\ll 1$, one could search among QNMs for QNM ratios which require large multiplicative factors to become integers and reduce or eliminate the degeneracy. We leave to future work the study of the ratios of QNMs. 

Concerning our resolution when determining $t_0$, typically in an experiment this is given by the period of the highest accessible frequency, since any error $\epsilon$ due to a finite SNR in estimating $t_0$ becomes more visible at higher modes:
\begin{align}
\textrm{Im(QNEC}(\omega)e^{i\omega(t_0+\epsilon)})\approx \textrm{QNEC}\,\omega\,\epsilon \,\, ,
\end{align}
where we set $t_0=0$ for convenience and expanded to $\mathcal{O}(\epsilon)\ll 1$. However since we expect QNEC$(\omega)$ decreases faster than $\omega^{-1}$ from the finiteness of the total GW energy, increasing $\omega$ does not yield the desired effect of enhancing the error to make it more visible. As discussed in Section \ref{sec:intro}, since the maximum SNR is fixed due to background fluctuations, we may in principle calculate a time resolution and share of temporal black hole information accessible in the signal, similarly to our estimate for the mass $m_0$. Finally, this argument also applies to the angular resolution of the black hole history, as we will see in the next section.

\section{Angle recovery}\label{sec:angle_recovery}
We now wish to recover the angle at which a particle fell radially into the black hole. We remind ourselves that the gravitational perturbations of a spherically symmetric background metric can be separated as such:
\begin{equation}
u(t,r,\theta,\phi)=\sum_{n=0}^\infty\sum_{l=2}^\infty\sum_{m=-l}^l\frac{u_{nlm}(r)}{r}Y^{lm}(\theta,\phi)e^{-i\omega_{nlm}(t+t_0)} \,\, ,
\end{equation}
where $u_{nlm}(r)$ is expressed in \ref{eq:solution_full} and we assume the integration half-circle radius $r_{\text{HC}}\rightarrow\infty$. This is because the full differential operator $\mathcal{L}=\mathcal{L}_{\text{Zerilli}}+\mathcal{L}_{\text{Legendre}}$ is separable into a radial and angular piece. We assume data from the GW wave signal of a radially infalling particle at $(\theta_1,\phi_1)$. We do not assume a reference frame with $\theta_1=0$ since we do not know $\theta_1$. Instead we assume an arbitrary reference frame $(\theta_0,\phi_0)$ at some fixed radius\footnote{We may think of the more complicated case without varying radius without loss of generality. We may also think of partial angular data of the GW. In this case we expect bandwidth considerations analogous to the time resolution (see previous section) to appear.}. A radially infalling particle will create only $m=0$ modes in a frame with $\theta_1=0$, since the differential equation $\mathcal{L}u=S$ is symmetric around the axis passing through the singularity and the particle and $\abs{m}>0$ modes are not. We may then write the solution~\ref{eq:signal}, assuming without loss of generality $r_{\text{HC}}\rightarrow\infty$ and writing $\omega_{nlm}\rightarrow\omega_{nl}$ for simplicity, in the form:
\begin{equation}
    u(t_0,r_0,\theta'+\theta_1,\phi'+\phi_1)=\sum_{n,l}u_{nl}(r_0)Y_{l0}(\theta',\phi')e^{i\omega_{nl}t_0} \,\, ,
\end{equation}
where $(\theta',\phi')$ are the coordinates in the $(\theta_1,\phi_1)$ frame, and unprimed coordinates mean we are in the $(\theta_0,\phi_0)$ frame. Note that while we express angles of $(\theta_0,\phi_0)$ and $(\theta_1,\phi_1)$ reference frames inside a single variable, these angles admit different representations in different frames which must be accounted for in computations, as we will see below. Nevertheless it may prove convenient to express the angles in various frames, and thus we use the prime/unprimed convention.

We now wish to recover $(\theta_1,\phi_1)$. To do so, we may solve (minimize) for $(\alpha,\beta)$ such that:
\begin{equation}\label{eq:harmonic_maximise}
    \int \sum_lY_{l0}(\theta',\phi')\sum_{l'}Y^*_{l'0}(\theta'+\alpha,\phi'+\beta) d\Omega
\end{equation}
is \textit{maximized} when integrated over the sphere, as we will see below. Note we dropped the sums over $n$ which factors out, and we wish to integrate $(\theta',\phi')$.
This allows us to maximize Equation \ref{eq:harmonic_maximise} for every $l\in \mathbb{N}$
because spherical harmonics are normalized ($\int Y Y^*d\Omega =1$) when $Y$ and $Y^*$ are identically oriented and centered. Here the second harmonics (unprimed) angles expressed in the $(\theta_0,\phi_0)$ reference frame are $(\theta'+\theta_2,\phi'+\phi_2)=(\theta'+\theta_1+\alpha,\phi'+\phi_1+\beta)$, where $(\theta_2,\phi_2)$ is the guess for the angle at which the particle is falling. Using Wigner's $D$-matrix we may write the expression as:
\begin{align} \nonumber
    & \int\sum_{l,l'} \sum_mD_{lm0}(\theta_1,\phi_1,0)Y_{lm}(\theta,\phi)\sum_{m'}D^*_{l'm'0}(\theta_1+\alpha,\phi_1+\beta,0)Y^*_{l'm'}(\theta,\phi)d\Omega\\ \nonumber
    = & \sum_{l,l'}\sum_{mm'}D_{lm0}(\theta_1,\phi_1)D^*_{lm'0}(\theta_1+\alpha,\phi_1+\beta,0)\int Y_{l0}(\theta,\phi)Y^*_{lm}(\theta,\phi)d\Omega\\ \nonumber
    = & \sum_{l,l'}\sum_{mm'}D_{lm0}(\theta_1,\phi_1)D^*_{lm'0}(\theta_1+\alpha,\phi_1+\beta,0)\delta_{l,l'}\delta_{m,m'} \\ \nonumber
    = & \sum_{l,m}D_{lm0}(\theta_1,\phi_1)\sum_nD^*_{lmn}(\theta_1,\phi_1)D^*_{ln0}(\theta',\phi',0) \\ \nonumber
    = & \, \delta_{n,0}\sum_{l,n}D^*_{ln0}(\theta',\phi',0) \\ = & \sum_l P_l(\cos\theta') \,\, ,
\end{align}
where we used $\sum_mD_{lm'm}(\theta_i,\phi_i)D_{lmn}(\theta'_j,\phi'_j)=D_{lm'n}(\theta_i+\theta_j,\phi_i+\phi_j)$ (closure of the D-matrix), 
$\sum_mD_{mm'}^*(\theta,\phi)D_{mn}(\theta,\phi)=\delta_{m',n}$ (orthogonality of the D-matrix) and $D_{l00}(\theta,\phi,0)=P_l(\cos\theta)$ where $P_l$ is the $l^\text{th}$ Legendre polynomial. The expression then gives
\begin{align}\label{eq:degeneracy_angle}\nonumber
    \sum_lP_l(\cos\theta'(\alpha,\beta,\theta_1,\phi_1)) & =\sum_lP_l(\cos\theta_1\cos(\theta_1+\alpha)+\sin\theta_1\sin(\theta_1+\alpha)\cos\beta) \\ \nonumber
    & \approx\sum_l P_l(\cos\alpha-\sin\theta_1\sin(\theta_1+\alpha)\beta^2/2)\\ \nonumber
    & \approx \sum_l P_l(1-\alpha^2/2-\beta^2\sin^2(\theta_1)/2)\\
    & \approx \sum_{l=l_\text{min}}^{l_\text{max}}(1-l(l+1)(\alpha^2+\beta^2\sin^2(\theta_1))/4) \,\, ,
\end{align}
where we used the spherical law of cosines in the first line then expanded to $\mathcal{O}(\alpha^2,\beta^2)$, and where $(l_{\text{min}},l_{\text{max}})$ are the lowest and highest angular resolution (i.e.~angular bandwidth) of the experiment. We see that, for any finite signal-to-noise ratio and $(\alpha,\beta)>0$, $l_{\text{max}}$ needs to be large enough to reject incorrect choice of $(\theta_2,\phi_2)$, and for an arbitrarily large SNR and $(\beta,\alpha)\ll 1$, we need arbitrarily large $l_{\text{max}}$. This is analogous to the determination of the time of infall $t_0$ for a single mode with $\omega_\text{max}\rightarrow l_\text{max}$.

Regarding angle periodic degeneracies in determining $(\theta_1,\phi_1)$, we note although it is $2\pi/l_{\text{min}}$ for $\abs{m}>0$, remarkably for $m=0$ no degeneracy is present. Finally, for multi-particle sources, preliminary separation of their signals is necessary before finding the angle of infall of each source and can be performed using highly-damped modes, as we will see in the next section.

\section{Resolving multiple particles and/or matter distributions}\label{sec:multiple_particles}
We have now determined the mass, time and angle of an infalling massive particle. However there remains the question whether two particles, or more generally a distribution of matter, can create degeneracies in the signal or can be distinguished.

Naively the sum of two two signals can be disentangled if one can solve for the masses and infall times of each particle $(m_0,m_1,t_0,t_1)$:
\begin{align}\label{eq:mode_sum} \nonumber
    A_{nl} & =e^{i\omega t_0}\left( m_0+m_1e^{i\omega(t_1-t_0)} \right) \\ \nonumber & =\frac{m_0e^{-\omega_It_0}\sin(\omega_R(t_1-t_0))}{\cos(\omega_Rt_1-\pi/2-\arctan(\frac{m_0e^{-\omega_It_0}\cos(\omega_R(t_1-t_0))+m_1e^{-\omega_It_1}}{m_0e^{-\omega_It_0}\sin(\omega_R(t_1-t_0))}))} \\ & \,\,\,  \times \exp(i\left(\omega_Rt_1-\pi/2-\arctan(\frac{m_0e^{-\omega_It_0}\cos(\omega_R(t_1-t_0))+m_1e^{-\omega_It_1}}{m_0e^{-\omega_It_0}\sin(\omega_R(t_1-t_0))})\right)) \, ,
\end{align}
where $A_{nl}$ represents \ref{eq:m0} (obtained from observation). In principle one may solve for the four variables using only two QNMs, thus forming four real equations (more generally, one needs $N$ QNMs to solve for the time and mass of $N$ sources), however the algebra is non-trivial as can be seen in the sum of two modes. Instead we may look at the set of QNMs to separate these equations using mode damping. Assuming an arbitrary desired accuracy in determining $m_0$, one may choose $\omega_{nl}$ with large enough $\omega_I$ (i.e.~overtone $n\gg 1$) such that, assuming $t_1>t_0$ without loss of generality, the term $m_1e^{i\omega_{nl}(t_1-t_0)}$ in the first line is suppressed by a factor of $e^{-\omega_{nl}^I(t_1-t_0)}$, where $\omega_{nl}^I$ is taken as positive, such that one can extract $m_0$ with an error of
$\mathcal{O}(m_1e^{-\omega_{nl}^I(t_1-t_0)})$. Indeed, for any $l$, $\omega_{nl}^R$ will converge to a constant while $\omega_{nl}^I$ diverges as $n\rightarrow\infty$, such that there is no limit to the arbitrary accuracy we demand. The closer the infall times $t_1$ and $t_0$ are, the larger $n$ will need to be to meet this criteria, and increasingly high $n$-modes will contain the information about increasingly smaller distances between particles. Assuming $N>2$ we may split the sources/particles into two groups of nearest neighbor particles such that all the particles of the first group fall in earlier than those in the second. We may then distinguish between the signal from each groups similarly to the case of two particles, and repeat the process of dividing each group into its components. Thus our reasoning generalizes to more complex structures, including classical objects formed of individual quanta, at which point a prescription for quantizing the gravitational effects would be necessary.

Looking at more general stress-energy tensors, this resolution should to apply to lightlike (massless) particles as well, although the corresponding signals are solved using different geodesics which we do not consider here, as long as the difference in mode amplitudes the potential provokes can be compensated by a choice of QNM mode with sufficiently large exponential suppression. Also note signals sourced by combinations of massive and massless particles (the most realistic sources) may be disentangled using this approach, with appropriate choices of QNMs, but is left for future work. Finally, considering particles with same infall times, but different infall angles, we may solve the angular analogue of \ref{eq:mode_sum}, calculated as sums of spherical harmonic contributions to the signal, to obtain the $N$ angles of infall, and associated amplitude of $N$ sources. Although we may not use the decay of modes as in the time domain to help us, numerical investigation has confirmed that the angular analogue of \ref{eq:mode_sum} admits a single solution at the correct values.

\section{Conclusion}
In this work we determined a complete classical information retrieval process using gravitational waves from a Schwarzschild black hole's history, and found that it is in principle possible to obtain to arbitrary accuracy in the absence of noise and with arbitrary computational resources. Indeed, from the GW frequencies one can recover the mass of the black hole, and the mode amplitudes contain the masses of the perturbers. Meanwhile, the infall times are contained in the phases of the amplitudes and can be separated using multiple modes, while the angle at which the particles fell in are contained in the angular phases of the GW signal modes. Therefore not only has all the information been recovered, but we have used every degree of freedom the signal contained contained.

Further, we may account for noise which is naturally present in the semiclassical regime as the black hole has a temperature. Nevertheless we show that the history of the mass of the black hole can be retrieved to $\mathcal{O}(1)$ when SNR$>1$ for granularity above the Planck scale (similarly the share of angular and temporal information can be determined). The share of \textit{quantum} information in GWs can therefore be quite large and can be estimated as the share of entanglement entropy between the black hole and GWs. Accounting for the measuring of the gravitational waves, the information recoverable in the black hole is then reduced from the initially large set of initial state possibilities with identical macroscopic number $M_\textrm{BH}$ to those consistent with the gravitational waves. 
Although the background fluctuations limiting our measurement have information as well, it is harder to retrieve/untangle and requires waiting until the end of the evaporation to obtain fully. This information is more akin to the ``closed system" information of the black hole studied in e.g.~\cite{Hayden:2007cs,Almheiri:2012rt,Dvali:2012rt}. Gravitational wave information is therefore a natural first place to look for information as it is more immediately available and relatively unentangled with other signals and the black hole.

We may put the importance of black hole information in gravitational waves in context by looking at three different regimes. In the purely classical regime ($\hbar\rightarrow 0$), black holes don't evaporate and information is considered lost behind the horizon and destroyed at the singularity. However initial state information is preserved in gravitational waves, where the information is copied and reflected at/near the horizon. 
In the semiclassical picture of quantized fields on a smooth metric background, black holes evaporate into maximally mixed states as shown by Hawking, containing no information besides the conservation of total energy of the initial state. In this context gravitational waves are the only source of information. Finally, in a purely quantum regime, one expects fluctuations of the black hole metric to contain information along with Hawking radiation. However a sizeable portion of the initial state information may still be in the gravitational waves emitted during the black hole history. Our finding of a Planck scale resolution is consistent with the expected quantum noise of gravitational waves below this scale, and implies Hawking radiation is induced by Planck scale fluctuations of the horizon.

Finally, looking ahead future work is needed in several directions, including finding the share of recoverable temporal and angular information, or generalizing our results to information recovery in rotating and electrically charged spacetimes. We may further wish to clarify the role of the branch cut modes, which cannot be described by QNMs, e.g.~whether they can describe a late time perturbative increase in mass of the black hole. Also our work is performed to first order in perturbation theory, in principle a self consistent description of the geodesic and perturbations is possible to arbitrary order. We hope that this work can stimulate further studies of gravitational waves in these directions.



\ack
LH would like to thank Ian Everbach, Wen-Yuan Ai, Lionel London, Oliver Lunt, Damian Galante, Sam Patrick and Joe Bhaseen for helpful discussions on gravitational waves, quantum many-body dynamics and the black hole information paradox, and Christopher McCabe for reading an early version of the manuscript. LH is supported by the King's Cromwell Scholarship. TT was supported by the STFC Quantum Technology Grants ST/T005858/1.

\appendix
\section{Source integral regularization}
We review how to regularize the divergent source term integral \ref{eq:source}. Expanding the integrand to first order in Schwarzschild coordinates with $2M=1$ and ignoring the boundary term as $t_0\rightarrow -\infty$:
\begin{equation}\label{eq:pre_Froebenius}
    I(\omega,r)u_1(\omega,r)\sim (r-1)^{1-2i\omega} \,\,\,\,\,\, r\rightarrow 1
\end{equation}
which diverges for all $2\omega_I<-1$. Following~\cite{PhysRevD.88.044018} we can rewrite the integrand as a Froebenius series:
\begin{equation}\label{eq:Froebenius}
    I(\omega,r)u_1(\omega,r)=\sum_n^\infty \xi_n(r-1)^{\zeta +n}
\end{equation}
where $\zeta=-2i\omega$ and following~\cite{PhysRevD.88.044018,1979ApJ...231..211D} we can regularize the integral by adding a boundary term (overall derivative):
\begin{equation}
    f(r)=\frac{d}{dr}\left(\sum_n^Nb_n\frac{(r-1)^{\zeta +n+1}}{\zeta +n+1}e^{-(r-1)}\right)
\end{equation}
where the integer $N=\max(n\leq -2\omega_I)_{n\in\mathbb{N}}$. The integral now converges if the $b_n$ coefficients are such that that only terms with positive real power remain in the integral. This corresponds to subtracting non-physical outgoing modes at the boundary $r\rightarrow 1$, and enforcing the horizon ingoing boundary condition that all integrand terms at the horizon which will contribute outgoing modes are set to zero. We thus match the series term by term to obtain the $(\xi_n,b_n)$ coefficients, and show here the first few terms here following~\cite{PhysRevD.88.044018}:
\begin{align}\nonumber
    b_0 & =\xi_0\\ \nonumber
    b_1 & =\xi_1+\frac{2+\zeta}{1+\zeta}b_0\\
    b_2 & =\xi_2+\frac{3+\zeta}{2(1+\zeta)}b_1-\frac{3+\zeta}{2(1+\zeta)}b_0
\end{align}
We see by comparison of \ref{eq:Froebenius} \& \ref{eq:pre_Froebenius} that $b_0=0$, $b_1=\xi_1$ and $b_n\propto\xi_1$ for $n\geq 1$.

\section*{References}

\bibliographystyle{iopart-num}
\bibliography{bibliography}

\end{document}